\documentclass[11pt]{article}
\usepackage[centertags]{amsmath}
\usepackage{amsfonts}
\usepackage{amssymb}
\usepackage{amsthm}
\usepackage{newlfont, times}

\usepackage[english]{babel}

\usepackage{epsfig, amssymb}
\newcommand{\finpr}{\hfill $\square$ \vspace{2mm}}

\newcommand{\bra}[1]{\langle#1|}
\newcommand{\ket}[1]{|#1\rangle}

\def\be{\begin{eqnarray}}
\def\ee{\end{eqnarray}}
\def\bee{\begin{eqnarray*}}
\def\eee{\end{eqnarray*}}
\newtheorem{thm}{Theorem}
\newtheorem{cor}{Corollary}
\newtheorem{lem}{Lemma}
\newtheorem{conj}{Conjecture}
\newtheorem{prop}{Proposition}

          \def\tr{\hbox{Tr}}

\begin{document}
\title{The LU-LC conjecture, diagonal local operations and quadratic
forms over GF(2)}

\author{David Gross\footnote{Institute for Mathematical Sciences, Imperial College
London, Prince's Gate, London SW7 2PG, UK.}
$^,$\footnote{QOLS, Blackett Laboratory, Imperial College
London, Prince Consort Road, London SW7 2BW, UK.}\quad and
Maarten Van den Nest\footnote{Institut f\"ur Quantenoptik
und Quanteninformation der \"Osterreichischen Akademie der
Wissenschaften, Innsbruck, Austria.}}

\date{\today}
\def\makeheadbox{}

\maketitle

\begin{abstract}
We report progress on the LU-LC conjecture---an open
problem in the context of entanglement in stabilizer states
(or graph states).  This conjecture states that every two
stabilizer states which are related by a local unitary
operation, must also be related by a local operation within
the Clifford group. The contribution of this paper is a
reduction of the LU-LC conjecture to a simpler
problem---which, however, remains to date unsolved. As our
main result, we show that, if the LU-LC conjecture could be
proved for the restricted case of \emph{diagonal} local
unitary operations, then the conjecture is correct in its
totality. Furthermore, the reduced version of the problem,
involving such diagonal local operations, is mapped to
questions regarding quadratic forms over the finite field
GF(2). Finally, we prove that correctness of the LU-LC
conjecture for stabilizer states implies a similar result
for the more general case of stabilizer codes.
\end{abstract}

\tableofcontents

\section{Introduction}

Stabilizer states---or, equivalently, graph states---are
special instances of multi-party quantum states that are of
interest in a number of domains in quantum information
theory and quantum computation. Stabilizer states are
defined in terms of the stabilizer formalism, which is a
group-theoretic framework originally designed in the 1990s
to construct broad classes of quantum error-correcting
codes---the \emph{stabilizer codes} \cite{Go97}. In
addition to their role in quantum error-correction, in
recent years stabilizer states have been considered in a
number of interesting applications, where the
measurement--based model of quantum computation known as
the \emph{one--way quantum computer} is certainly among the
most prominent \cite{Ra01, Ra03}. We refer to Ref.
\cite{He06} for a recent overview article about stabilizer
states and their applications.

It is well known that many stabilizer states exhibit a high
degree of genuine multi-party entanglement \cite{He06}, and
that this entanglement is a key ingredient responsible for
the successful use of these states in various applications.
Therefore, a detailed study of the entanglement properties
of stabilizer states is of natural interest. Recently, a
number of authors have studied this topic with considerable
success (for an incomplete list, see Refs. \cite{He06,He04,
Du04, Fa04, Va04a, Va04b, He05,  Gr05, Va05, Da05,
Br06, Gr06, Ze06, Gr07, Ma07}), and also the present paper is
situated in this context.

The study of the nonlocal properties of stabilizer states
naturally leads to an investigation of the action of local
unitary (LU) operations on stabilizer states, and a
classification of stabilizer states under LU equivalence.
In this context, an important role is played by a subclass
of LU operations known as \emph{local Clifford} (LC)
operations, which are defined to be those LU operations
mapping the Pauli group to itself under conjugation. Due to
the close connection between the Pauli group, the
stabilizer formalism and the (local) Clifford group, the
action of LC operations on stabilizer states can be
described efficiently and in a transparent manner, allowing
for a thorough understanding of the entanglement in
stabilizer states with respect to this restricted LC
symmetry. For example, the action of LC operations on graph
states can entirely be understood in terms of a single
elementary graph transformation rule \cite{Va04a}.
Moreover, a systematic classification of LC equivalence of
stabilizer states is possible and has been executed up to
$n=12$ qubits \cite{He04, Da05}. Finally,  an efficient
algorithm (i.e., with polynomial time complexity in the
number of qubits) to decide whether two given stabilizer
states are LC equivalent, is known \cite{Va04b}.

In the study of LU equivalence of stabilizer states, it is
natural to ask whether the restriction to LC equivalence is
in fact a restriction at all. This is the content of the
``LU-LC conjecture'', which states that \emph{``Every two
LU equivalent stabilizer states must also be LC
equivalent''}. The conjecture, which will be the central
topic of this paper, has been listed as the 28th open
problem in quantum information theory \cite{openprob}. The
main implication of a proof of the LU-LC conjecture would
be that questions regarding entanglement of stabilizer
states can entirely be treated within the closed framework
of stabilizer formalism plus local Clifford group. In
particular, the aforementioned insights into the restricted
regime of  LC equivalence would then count as insights
regarding the ``true'' local unitary symmetry. Even more
so, in previous work it was shown that the notions of LU
equivalence and equivalence under stochastic local
operations and classical communication (in short:
\emph{SLOCC equivalence}) coincide for all stabilizer
states \cite{Va04c}. Therefore, correctness of the LU-LC
conjecture would imply that both of these symmetries would
be reduced to the tractable case of LC equivalence.

The LU-LC conjecture has been studied considerably in
recent years. The most recent progress involved proofs that
LU and LC equivalence indeed coincide for large subclasses
of stabilizer states \cite{Va05, Ze06}, but a complete
proof of the conjecture remained---and remains---out of
reach. In this work, we report further significant
advances. Because the argument will be technical, at this
point we give a brief outline of the results.

The pivotal conjecture is the following:

\begin{conj}\label{conj_LULC_codes}{\bf (LU-LC conjecture)}
Every two LU equivalent stabilizer states are also LC
equivalent.
\end{conj}
\noindent It is the aim of this work to reduce the
LU-LC conjecture to a simpler problem\footnote{Note that
this approach is different from the one adopted in e.g.
Refs. \cite{Va05, Ze06}, where one aims at constructing
as-large-as-possible subclasses of stabilizer stats for
which the conjecture holds.
}. This reduction will take place in a number of steps.
First, a central finding in the present paper will be that
only a very restricted class of LU operations has the
capability of mapping a stabilizer state to another
stabilizer state. The following theorem was first proved in
one of the authors' diploma thesis \cite{Gr05}.  In this paper, we
present a more direct argument.
[After this work had been completed, B.\ Zeng pointed out to us that
the same statement had been obtained independently in Ref.\ \cite{Ze07}.]

\begin{thm}{\bf (Reduction to diagonal unitaries)}
    \label{cor:diagonalsIntro}
    Let $U=U_1\otimes\dots\otimes U_n$ be an LU operation, and suppose that there exist stabilizer states
    $|\psi\rangle$ and $|\psi'\rangle$ such that $U|\psi\rangle =
    |\psi'\rangle$.  Then, up to the action of Clifford operations,
        all $U_i$ are diagonal matrices.

        More precisely,
        every one-qubit operator $U_i$ has the form $U_i = C_i
    D_iC_i'$, where $C_i$ and $C_i'$ are Clifford and $D_i$ is diagonal.
\end{thm}
This result immediately implies that the LU-LC conjecture
is equivalent to the following simpler problem: ``Every two
stabilizer states that are related by a \emph{diagonal} LU
operator, are also LC equivalent''. This provides the first
reduction of the LU-LC conjecture: only diagonal LU
operations need to be considered. Note that a diagonal
unitary operator on a single qubit has the form
$\operatorname{diag}(1,e^{i \phi})$ and therefore depends
only on one real parameter. This is a significant reduction
in complexity w.r.t. to the case of general $SU(2)$
operators, which depend on three parameters. Section
\ref{sec:fromLUtoDLU} is devoted to the proof of Theorem
\ref{cor:diagonalsIntro}.

In Section \ref{sec:quadraticForms}, we will show that the
remaining problem is related to a certain statement about
quadratic forms and linear spaces over the finite field of
order two (which will be denoted by GF(2) or, equivalently,
$\mathbb{F}_2$). The following result will be obtained.

\begin{thm}{\bf (Reduction to quadratic forms)}
    \label{thm:quadraticForms}
    Let $S$ be a linear subspace of $\mathbb{F}_2^n$, and let
    $Q:\mathbb{F}_2^n\to\mathbb{F}_2$ be a quadratic function. Suppose
    that there exist complex phases $\{c_i\}$, such that
    \be\label{eq_thm:quadraticForms}
        (-1)^{Q(x)} = \prod_{i=1}^n c_i^{x_i},\quad  \mbox{for every $x\in S$}.
    \ee
    If, for every such $Q$ and $S$, the phases can always be chosen from $\{\pm1, \pm i\}$, then the
    LU-LC conjecture is true.
\end{thm}

The criterion in the preceding theorem is not only
sufficient for the LU-LC conjecture, but---up to a sensible
extra assumption---also necessary. Hence, essentially, the
pertinent question reduces to a problem concerning binary
quadratic forms---note that there is no mentioning about
stabilizer states or local unitary operations in the
formulation (\ref{eq_thm:quadraticForms}). Remarkably, the
LU-LC problem remains hard even in this considerably
simplified guise, and a proof (or counterexample) has to
date not been found.

As a final result in this paper, in Section
\ref{sec:fromCodesToStates} we will prove that correctness
of the LU-LC conjecture for stabilizer states would imply a
similar LU-LC theorem regarding the more general  case
of stabilizer \emph{codes}---recall that stabilizer states
are a specific instance of stabilizer codes (they form the
class of one-dimensional codes). The following result will
be proven:

\begin{thm}\label{thm:statesSuffice}
    {\bf (Reduction from codes to states)}
    The LU-LC conjecture holds for all stabilizer codes if and only if
    it holds for stabilizer states.
\end{thm}

Therefore, in conjunction with theorems 1 and 2, this
result implies that the general LU-LC conjecture for
stabilizer codes is reduced to the problem regarding
quadratic forms over GF(2) as posed in Eq.
(\ref{eq_thm:quadraticForms}).

\section{Stabilizer states and codes, and local equivalence}

In this section we fix some notations, state basic definitions, and
recall some preliminary results which will be needed in the following.
For more details, we refer the reader to Refs. \cite{Go97, QCQI}.

\subsection{Stabilizer states and codes}
\label{sec:stabs}

The $2^n\times 2^n$ identity matrix is denoted by $I_n$,
for every $n\in\mathbb{N}_0$. The $n$-qubit Hilbert space
is ${\mathcal H}_n\cong \mathbb{C}^{2^n}$.

The Pauli group ${\mathcal G}_1$ on one qubit is the
multiplicative subgroup of $U(2)$ generated by the Pauli
matrices \be X = \left[\begin{array}{cc}
0&1\\1&0\end{array} \right],\ Y = \left[\begin{array}{cc}
0&-i\\i&0\end{array} \right],\ Z = \left[\begin{array}{cc}
1&0\\0&-1\end{array} \right].\ee Note that the Pauli
matrices $X$, $Y$ and $Z$ are Hermitian and unitary
operators with zero trace. The Pauli group ${\mathcal G}_n$
on $n$ qubits is the $n$-fold tensor product of ${\mathcal
G}_1$ with itself. For an arbitrary $n$-qubit Pauli
operator $g\in{\mathcal G}_n$, we let $g_1, \dots,
g_n\in\{I, X, Y, Z\}$, written with \emph{lower} indices,
denote the unique one-qubit Pauli operators such that $g
\propto g_1\otimes\dots\otimes g_n$. Here $\propto$ denotes
equality up to a global phase factor.
%
%
%
The \emph{support} of an $n$-qubit Pauli operator $g$ is
the set \be \mbox{supp}(g)=\{i\in\{1, \dots, n \}\ |\
g_i\neq I_1\}.\ee The operator $g$ is said to have
\emph{full support} if supp$(g)=\{1, \dots, n\}$.
We will use the shorthand notations
\be\label{eqn:shortXZ}
    Z(t):=\bigotimes_{i=1}^n Z^{t_i}
    \quad\mbox{and}\quad
    X(t):=\bigotimes_{i=1}^n X^{t_i},
\ee
for every $t=(t_1, \dots, t_n)\in\{0,1\}^n$.

A \emph{stabilizer} ${\mathcal S}$ on $n$ qubits is defined
to be an Abelian subgroup of ${\mathcal G}_n$ that does not
contain $-I$. The following is a list of elementary
properties of stabilizers, which can be found in the
literature \cite{Go97, QCQI}.

\begin{itemize}
\item Every element $g$ of a stabilizer ${\mathcal S}$ has
the form $g = \pm g_1\otimes \dots\otimes g_n$, where
$g_i\in\{I, X, Y, Z\}$. It follows that stabilizer elements
are always both Hermitian and unitary operators. In
particular, one has $g^2=I_n$. \item If $g\in {\mathcal S}$
then $-g\notin {\mathcal S}$. \item The trace of a
stabilizer element different from the identity is equal to
zero. \item The cardinality $|{\mathcal S}|$ of the
stabilizer ${\mathcal S}$ is always a power of two not
greater than $2^n$. If $|{\mathcal S}|=2^k$ then ${\mathcal S}$ is generated by $k$ independent elements. The number $k$ is then called the \emph{rank} of ${\mathcal S}$.

\end{itemize}

The \emph{stabilizer code} associated to an $n$-qubit
stabilizer ${\mathcal S}$ is the subspace $V_{\mathcal
S}\subseteq {\mathcal H}_n$ consisting of all simultaneous
fixed points of the elements of ${\mathcal S}$, i.e., \be
V_{\mathcal S} :=\{|\psi\rangle\in {\mathcal H}_n\ |\
g|\psi\rangle = |\psi\rangle\mbox{ for every } g\in
{\mathcal S}\}.\ee The dimension of $V_{\mathcal S}$ is
equal to $2^n|{\mathcal S}|^{-1}$, which is a power of two.
The stabilizer code $V_{\mathcal S}$ is identified with the
operator \be\label{rho} \rho := \frac{1}{2^n}
\sum_{g\in{\mathcal S}} g,\ee which is, up to a
multiplicative constant, equal to the orthogonal projector
on the code $V_{\mathcal S}$. The normalization is chosen
such as to yield Tr$(\rho)=1$.

If ${\mathcal S}$ is an $n$-qubit stabilizer with
cardinality $|{\mathcal S}|=2^n$, the code $V_{\mathcal S}$
is one-dimensional, or, equivalently, the associated
projector $\rho$ has rank one and is therefore of the form
\be\rho = |\psi\rangle\langle\psi|\ee for some
$|\psi\rangle\in{\mathcal H}_n$. The class of pure states
$|\psi\rangle$ that are obtained in this way are called
\emph{stabilizer states}. Thus, a stabilizer state on $n$
qubits is any state $|\psi\rangle$ having the property that
$g|\psi\rangle = |\psi\rangle$ for every element $g$ in a
maximal stabilizer ${\mathcal S}$, i.e., where $|{\mathcal
S}|=2^n$. We refer to Ref. \cite{He06} for a recent review of
stabilizer states and their properties.

\subsection{Local equivalence}

We now introduce the notions of local equivalence of
stabilizer states and codes that we will study in the
following.

{\it LU equivalence.---} Two stabilizer codes\footnote{In
this and the following definitions in this section, we
consider stabilizer states as one-dimensional instances of
stabilizer codes,  $\rho = |\psi\rangle\langle\psi|$.}
$\rho$ and $\rho'$ are called LU equivalent if there exists
a local unitary operator $U\in U(2)^{\otimes n}$ such that
$U\rho U^{\dagger}=\rho'$.

{\it LC equivalence.---} A $2\times 2$ unitary operator $U$
is called a \emph{Clifford} operator\footnote{
    Note that the group of Clifford operators which appears in quantum
    information theory has nothing to do with either Clifford algebras
    or the Clifford group used e.g.\ in the context of Fermionic systems
    or the representation theory of $SO(n)$.
} on one qubit if $U\sigma U^{\dagger}\in{\mathcal G}_1$
for every Pauli matrix $\sigma\in\{X, Y, Z\}$. The set of
all Clifford operations forms a matrix group called the
Clifford group. It can be shown that the Clifford group is
generated by the the matrices \be c
I_1,\quad \frac{1}{\sqrt{2}} \left[\begin{array}{cc}1 & 1\\
1 & -1
\end{array}\right] \mbox {\quad and \quad} \left[\begin{array}{cc}1 & 0\\ 0 & i
\end{array}\right],\ee where $c$ ranges over all complex phases.
Note that the Pauli matrices $X$, $Y$ and $Z$
are instances of Clifford operations.  A \emph{local
Clifford} operator (LC operator) on $n$ qubits is a local
unitary operator $U = U_1\otimes\dots\otimes U_n$, where
every tensor factor $U_i$ is a Clifford operator. Two
stabilizer codes are called LC equivalent if there exists
an LC operator $U$ relating the two codes under
conjugation.

{\it Semi-Clifford operations.---} An important ingredient
in the following will be a third kind of local operations,
namely the \emph{local semi-Clifford operations}, which are
defined next. A $2\times 2$ unitary operator $U$ is called
a \emph{semi-Clifford} operator on one qubit if there exist
a Pauli matrix $\sigma\in\{X, Y, Z\}$ such that $U\sigma
U^{\dagger}\in{\mathcal G}_1$. Thus, a semi-Clifford
operator is defined to send \emph{at least one} of the
Pauli matrices to another Pauli matrix under conjugation
(up to a global phase factor). As an example, the diagonal
matrix \be D=\left[\begin{array}{cc}1 & 0\\ 0 & c
\end{array}\right],\ee where $c$ is an arbitrary complex phase, is a semi-Clifford
operator for all $c$, since $DZD^{\dagger}=Z$. However, $D$
is only a Clifford operation if $c\in\{\pm 1, \pm i\}$. It
is clear that every Clifford operator is also a
semi-Clifford. We then define a \emph{local semi-Clifford}
operator on $n$ qubits to be a local unitary operator $U =
U_1\otimes\dots\otimes U_n$, where every tensor factor
$U_i$ is a semi-Clifford operator.

\section{From LU to diagonal LU operations}
\label{sec:fromLUtoDLU}

In this section we show that there exist severe
restrictions on the LU operators which can realize local
transformations between stabilizer codes (or states). In
particular, we will prove that \emph{any LU operator
mapping a stabilizer code (or state) to another one must be
a semi-Clifford operation}.  We will subsequently use this
result to show that, in the study of the LU-LC conjecture,
one can---without loss of generality---restrict attention
to local equivalence of stabilizer states and codes with
respect to \emph{diagonal} LU operations only, i.e., LU
operations of the form \be U = c\cdot
\left[\begin{array}{cc} 1 & \\ & c_1\end{array}
\right]\otimes\dots\otimes \left[\begin{array}{cc} 1 & \\ &
c_n\end{array} \right],\ee where $c, c_1, \dots, c_n$ are
complex phases. Hence, the complexity of the LU operations
which need to be considered in the study of the LU-LC
conjecture is drastically reduced.

In Section \ref{sect_LU_to_DLU_prelim} some preliminary
results are proven. In Section \ref{sect_LU_to_DLU_semi} we
show that any LU operator mapping a stabilizer code (or
state) to another one is necessarily a semi-Clifford
operation. Finally, in Section \ref{sect_LU_to_DLU_final}
we show that this allows one to restrict attention to
diagonal LU operations in the study of the LU-LC
conjecture.

\subsection{Preliminary results}\label{sect_LU_to_DLU_prelim}

Below, the following type of stabilizer codes will play a
role. Let $m\in \mathbb{N}_0$. A $[2m, 2m-2, 2]$ stabilizer
code is a code with stabilizer of the form \be {\mathcal
S}= \{I_{2m},\ g,\ g',\ gg'\},\ee where $g,\ g'$ and $gg'$
are Pauli operators having full support. Every $[2m, 2m-2,
2]$ code is LU equivalent
to the code $\rho^{[2m, 2m-2, 2]}$
defined by \be \rho^{[2m, 2m-2, 2]}:=\frac{1}{4^m} (I_{2m}
+ X^{\otimes 2m} + (-1)^m Y^{\otimes 2m} + Z^{\otimes
2m}).\ee The operator $\rho^{[2, 0, 2]}$
has rank one, and is therefore a stabilizer state.
Concretely, one has $\rho^{[2, 0, 2]} =
|\psi^+\rangle\langle\psi^+|,$ where $|\psi^+\rangle =
\frac{1}{\sqrt{2}}(|00\rangle + |11\rangle)$ is the EPR
state. The following result was proven in Ref.
\cite{RainsMin2} and will be an important part of our
analysis.

\begin{prop}\cite{RainsMin2}\label{prop_rains}
Let $m\in\mathbb{N}_0$, $m\geq 2$. Let $\rho$ and $\rho'$ be two
$[2m, 2m-2, 2]$ stabilizer codes and let $U\in U(2)^{\otimes n}$
be an LU operator such that $U\rho U^{\dagger}=\rho'$. Then $U$ is
an LC operator.
\end{prop}

For every subgroup ${\mathcal T}$ of ${\mathcal S}$, the
\emph{index} of ${\mathcal T}$ in ${\mathcal S}$ is defined
to be the number $[{\mathcal S}: {\mathcal T}]:= |{\mathcal
S}||{\mathcal T}|^{-1}.$ Note that $|{\mathcal S}|$ is a
power of two, and therefore $|{\mathcal T}|$ and
$[{\mathcal S}: {\mathcal T}]$ are also powers of two. For
every $i= 1,\dots, n$, define ${\mathcal S}\langle i\rangle
:= \{g\in {\mathcal S}\ |\ g_i = I_1\}.$ It is easily
verified that ${\mathcal S}\langle i\rangle$ is a subgroup
of ${\mathcal S}$. We will need the following lemmas.

\begin{lem}\label{lem1}
Let ${\mathcal S}$ be a stabilizer on $n$ qubits. Then
$[{\mathcal S}: {\mathcal S}\langle i\rangle]\in\{1, 2,
4\},$ for every $i=1, \dots, n$.
\end{lem}
{\it Proof:} the proof uses elementary group theory. We
start from the property that ${\mathcal S}$ can be
partitioned into cosets of the subgroup ${\mathcal
S}\langle i\rangle$:\be {\mathcal S}=  g^{(1)} {\mathcal
S}\langle i\rangle \cup\dots\cup g^{(N)} {\mathcal
S}\langle i\rangle, \ee for some Pauli operators
$g^{(1)}=I_n,\ g^{(2)}, \dots, g^{(N)}\in{\mathcal S}$,
where \be g^{(j)} {\mathcal S}\langle i\rangle\cap g^{(k)}
{\mathcal S}\langle i\rangle=\emptyset\ee for every $j,
k=1, \dots, N$ with $j\neq k$. The number of cosets $N$ is
equal to $[{\mathcal S}: {\mathcal S}\langle i\rangle]$.
Note that two elements $g, g'\in{\mathcal S}$ belong to
different cosets of ${\mathcal S}\langle i\rangle$ if and
only if $g_i\neq g'_i$, showing that there can be at most 4
cosets, as $g_i\in\{I_1, X, Y, Z\}$. Since $[{\mathcal S}:
{\mathcal S}\langle i\rangle]$ is a power of two, the
result follows. \hfill $\square$

\begin{lem}\label{lem:luInvariants}
Let $\rho$ be an $n$-qubit  stabilizer  code with
stabilizer ${\mathcal S}$, and let $i\in\{1, \dots, n\}$.
Then the quantities $|{\mathcal S}|$, $|{\mathcal S}\langle
i\rangle|$, and $[{\mathcal S}': {\mathcal S}'\langle
i\rangle]$ are local unitary invariants.
\end{lem}
{\it Proof: } we have seen in Section  \ref{sec:stabs} that the rank of
$\rho$ is equal to $2^n |{\mathcal S}|^{-1}$. As the rank
of a density operator is an LU invariant, this shows that
$|{\mathcal S}|$ is an LU invariant. Second, it was proven
in Ref.\ \cite{Va05}  that the quantities $|{\mathcal S}\langle
i\rangle|$ are LU invariants. It then immediately follows
that the quantities $[{\mathcal S}': {\mathcal S}'\langle
i\rangle]$ are LU invariants as well. \finpr

\begin{lem}\label{lem2}
Let  $\rho$ and $\rho'$ be LU equivalent stabilizer codes with
stabilizers ${\mathcal S}$ and ${\mathcal S}'$, respectively. Let
$U=U_1\otimes\dots\otimes U_n\in U(2)^{\otimes n}$ such that
$U\rho U^{\dagger}=\rho'$. Then $U_i$ is semi-Clifford for every
$i\in\{1, \dots, n\}$ for which $[{\mathcal S}: {\mathcal S}\langle
i\rangle]=2$.
\end{lem}
{\it Proof:} Let $i\in\{1, \dots, n\}$ such that $[{\mathcal S}: {\mathcal
 S}\langle i\rangle]=2$. Since  $\rho$ and $\rho'$ are locally equivalent,
we also have $[{\mathcal S}': {\mathcal S}'\langle
i\rangle]=2$ from Lemma \ref{lem:luInvariants}. Therefore,
we can partition ${\mathcal S}$ and ${\mathcal S}'$ in
cosets as follows: ${\mathcal S} = {\mathcal S}\langle
i\rangle\cup g{\mathcal S}\langle i\rangle$ and ${\mathcal
S}' = {\mathcal S}'\langle i\rangle\cup g'{\mathcal
S}'\langle i\rangle,$ where $g\in{\mathcal
S}\setminus{\mathcal S}\langle i\rangle$ and
$g'\in{\mathcal S}'\setminus{\mathcal S}'\langle i\rangle$.
Defining $ \rho\langle i\rangle= \frac{1}{2^n}
\sum_{h\in{\mathcal S}\langle i\rangle} h $ and
$\rho'\langle i\rangle$ similarly, it follows from the
definitions of $\rho$ and $\rho'$  that
\be\label{index}\rho = (I_n+g)\rho\langle
i\rangle\quad\mbox{and} \quad \rho' = (I_n+g')\rho'\langle
i\rangle.\ee Note that \be\label{partialtrace} \rho\langle
i\rangle = \mbox{Tr}_i(\rho)\otimes \frac{I_1}{2} \mbox{
and } \rho'\langle i\rangle = \mbox{Tr}_i(\rho')\otimes
\frac{I_1}{2}.\ee This property essentially follows from
the fact that, in taking the partial trace over the $i$th
qubit, the only Pauli operators in the expansion
(\ref{rho}) which survive the partial trace are those
having an $i$th tensor factor equal to the identity.

Using the identity $U\rho U^{\dagger}=\rho'$ and
(\ref{partialtrace}), we have $U\rho\langle i\rangle
U^{\dagger}=\rho'\langle i\rangle$. It then follows from
(\ref{index}) that $\left(UgU^{\dagger}\right) \rho'\langle
i\rangle = g'\rho'\langle i\rangle.$ The r.h.s. of this
equation is a sum of Pauli operators all having the same
$i$th tensor factor, namely $g_i'$. Therefore, the l.h.s.
must also have this property, and this can only occur if
$U_ig_iU_i^{\dagger}\propto g_i'$. Since $g_i\neq I_1\neq
g_i'$, this shows that $U_i$ is semi-Clifford. \hfill
$\square$

\begin{lem} \label{lem3} Let ${\mathcal S}$ be a stabilizer on $n$ qubits and let $\Pi$
be the smallest subgroup of ${\mathcal S}$ containing all
subgroups ${\mathcal S}\langle i\rangle$, i.e., \be \Pi=
\left\{ g^{(1)}g^{(2)}\dots g^{(n)} |\ g^{(i)}\in{\mathcal
S}\langle i\rangle,\ i=1, \dots, n\right\}.\ee Then one of
the following three cases occurs:
\begin{itemize} \item[(i)] $\Pi = {\mathcal S}$; \item[(ii)] $[{\mathcal S}:\Pi]=
 2$; \item[(iii)] $[{\mathcal S}:\Pi]= 4$; in this case, the associated code must be a $[2m, 2m-2, 2]$ code.
\end{itemize}
\end{lem}
{\it Proof}: Since $\Pi$ is a subgroup of ${\mathcal S}$, $[{\mathcal
S}:\Pi]$ is a power of two. Furthermore, each ${\mathcal S}\langle
i\rangle$ is a subgroup of $\Pi$ and therefore $[{\mathcal S}:\Pi]
\leq [{\mathcal S}: {\mathcal
 S}\langle i\rangle]\leq 4$, for every $i=1, \dots, n$.  This shows that
 $[{\mathcal S}:\Pi]\in\{1,2,4\}.$  We investigate
these possibilities case by case. First, if $[{\mathcal S}:\Pi]= 1$
then $\Pi={\mathcal S}$ trivially, which proves (i).

We now prove (iii). If $[{\mathcal S}:\Pi]= 4$ then  ${\mathcal S}$ can be
partitioned in cosets as follows:
\begin{equation}
    {\mathcal S} = \Pi \cup g^{(1)}\Pi \cup g^{(2)}\Pi\cup g^{(3)}\Pi,
\end{equation}
for suitable $g^{(j)}\in {\mathcal S} \setminus \Pi$. The
$g^{(j)}$ must have full support and must pairwise differ
on every qubit. For, suppose there is a qubit $i$ such
that, say, $g^{(1)}_i = g^{(2)}_i$. Then $g^{(1)}
g^{(2)}\in \Pi$, implying that $g^{(1)}\Pi=g^{(2)}\Pi$,
which contradicts the definition of the $g^{(j)}$. A
similar argument can be given for arbitrary pairs $g^{(j)}$
and $g^{(k)}$. This shows that the $g^{(j)}$s must pairwise
differ on every qubit.

Next, let $f$ be an arbitrary element of $\Pi$. We prove
that $f$ must be equal to the identity by contradiction:
suppose there is a qubit $i$ such that $f_i\neq I_1$, then
there exists a $j\in\{1, 2, 3\}$ such that $f_i=g^{(j)}_i$. But
this implies that
\begin{equation}
    g^{(j)} = \underbrace{\vphantom{(}f}_{\in \Pi}\  \underbrace{(f g^{(j)})}_{\in \Pi} \in
    \Pi
\end{equation}
which is a contradiction. Hence $f=I_n$, so $\Pi=\{I_n\}$
and $|S|=4$. But then $S=\{g^{(1)},g^{(2)},g^{(3)},I_n\}$,
proving the claim. \finpr

\subsection{Semi-Clifford operations}\label{sect_LU_to_DLU_semi}

We are now in a position to prove the main results of this
section. Defining the \emph{support}\footnote{This
definition is introduced for technical reasons. If the
support of a stabilizer on $n$ qubits is strictly contained
within the set $\{1, \dots, n\}$, then the associated code
can be written as the product of a code on fewer qubits and
the identity matrix. Therefore, for any reasonable
application it makes no sense to consider stabilizers not
having full support. This definition is however introduced
here to facilitate the induction argument made in the proof
of Theorem \ref{partial}.} of a stabilizer ${\mathcal S}$
to be the set $\mbox{supp}({\mathcal S}):=
\bigcup_{g\in{\mathcal S}} \mbox{ supp}(g),$ we can
precisely  formulate the main result of this section.

\begin{thm}\label{partial}
Let $\rho$ and $\rho'$ be LU equivalent stabilizer codes
with stabilizers ${\mathcal S}$ and ${\mathcal S}'$ on
$n\geq 2$ qubits, and suppose that $\rho$ cannot be written
as a product of the form \be\label{requirement}
|\psi\rangle\langle\psi|\otimes \rho'',\ee where
$|\psi\rangle$ is a 2--qubit stabilizer state LU equivalent
to the EPR state and $\rho''$ is a stabilizer code on $n-2$
qubits. Let $U=U_1\otimes \dots\otimes U_n\in U(2)^{\otimes
n}$ such that $U\rho U^{\dagger} = \rho'$. Then $U_i$ is
semi-Clifford for every $i\in\mbox{supp}({\mathcal S})$.
\end{thm}
{\it Proof:} We prove the result
by induction on $n$. If $n=2$, up to local equivalence plus
permutations of the 2 qubits the following stabilizer codes
$\rho$ fulfilling the requirement of the theorem exist: \be
4\rho = \left \{
\begin{array}{l} I_2\\I_2+ Z\otimes Z\\ I_2+ I_1\otimes Z\\ I_2+
I_1\otimes Z+ Z\otimes I_1+ Z\otimes Z
\end{array}\right..\ee It is straightforward to verify that the claim holds for these
codes.

In the induction step of the proof, fix $n\geq 3$ and suppose the
result has been verified for all $n'< n$. Let $\rho$ and $\rho'$ be locally
equivalent stabilizer codes on $n\geq 3$ qubits satisfying the
requirement of the theorem, and let $U=U_1\otimes\dots\otimes
U_n\in U(2)^{\otimes n}$ such that $U\rho U^{\dagger} = \rho'$. It
follows that \be U[i]\mbox{ Tr}_i(\rho) U[i]^{\dagger} = \mbox{
Tr}_i(\rho')\ee for every $i=1, \dots, n$, where we have defined
\be U[i] := \ U_1\otimes\dots\otimes U_{i-1}\otimes
U_{i+1}\otimes\dots\otimes U_n.\ee Note that $\mbox{ Tr}_i(\rho)$
and $\mbox{ Tr}_i(\rho')$ are stabilizer codes on $n-1$ qubits,
and that $\mbox{ Tr}_i(\rho)$ cannot be written as a product
in the form of (\ref{requirement}).  We can therefore apply the induction
hypotheses to every pair $\mbox{ Tr}_i(\rho)$ and $\mbox{
Tr}_i(\rho')$, where $i=1, \dots, n$. This proves that $U_j$ is
semi-Clifford for every $j$ in the set \be\label{supp}
\bigcup_{i=1}^n \mbox{ supp } {\mathcal S}\langle i\rangle.\ee Now, if
the set (\ref{supp}) is equal to $\mbox{ supp}({\mathcal S})$ then we
are done. If this is not the case, then there exist $j\in\mbox{
supp}({\mathcal S})$ such that $j\notin \mbox{ supp}({\mathcal S}\langle
i\rangle)$ for every $i=1, \dots, n$, and hence $j\notin \mbox{
supp}(\Pi)$, where $\Pi$ is defined as in Lemma \ref{lem3}. This
last property implies that $\Pi\neq {\mathcal S}$, and therefore case
(ii) or case (iii) in Lemma \ref{lem3} must apply.

If case (ii) holds, the stabilizer ${\mathcal S}$ can be
written as a partition \be\label{ii}{\mathcal S}=\Pi \cup g
\Pi,\ee where $g\in{\mathcal S}\setminus\Pi$, and therefore
$g$ has full support. Expression (\ref{ii}) together with
the property that $j\notin \mbox{ supp }(\Pi)$ implies that
$h_j\in\{I_1, g_j\}$ for every $h\in {\mathcal S}$, and
thus $[{\mathcal S}: {\mathcal S}\langle j\rangle]=2$.
Lemma \ref{lem2} then shows that $U_j$ must be a
semi-Clifford operation.

In the event of case (iii), $\rho$ and $\rho'$ must be $[2m, 2m-2,
2]$ codes with $m\neq 1$, and proposition \ref{prop_rains} then
implies that $U$ is a local Clifford operation, which is a
fortiori local semi-Clifford. This proves the result. \hfill
$\square$

As an immediate corollary of this result, we find:

\begin{cor}\label{cor1}
Let $|\psi\rangle$ and $|\psi'\rangle$ be fully entangled,
LU equivalent stabilizer states on $n\geq 3$ qubits, and
let $U\in U(2)^{\otimes n}$ be an LU operator such that
$U|\psi\rangle =|\psi'\rangle$. Then $U$ is a local
semi-Clifford operator.
\end{cor}
{\it Proof:} letting ${\mathcal S}$
be the stabilizer of $|\psi\rangle$, it is clear that
${\mathcal S}$ has full support. Moreover, $|\psi\rangle$
is a fully entangled state on $n\geq 3$ qubits and
therefore satisfies the requirements of Theorem
\ref{partial}. The result follows immediately.\finpr

From this point on, we will only consider fully entangled
stabilizer states on $n\geq 3$ qubits. Note that the
restriction to fully entangled states does not entail a
loss of generality.

\subsection{Diagonal LU operations}\label{sect_LU_to_DLU_final}

Let $|\psi\rangle$ and $|\psi'\rangle$ be stabilizer states
on $n$ qubits and let $U=U_1\otimes\dots\otimes U_n$ be an
LU operator such that $U|\psi\rangle=|\psi'\rangle$.
According to corollary \ref{cor1}, $U$ must be a local
semi-Clifford operation. By definition, this means that there exist $n$
Pauli matrices $\sigma_i\in\{X, Y, Z\}$ such that $
U_i\sigma_i U_i^{\dagger}\in{\mathcal G}_1$ for every $i=1,
\dots, n$. It is then easy to verify that there exist LC
operators $V=V_1\otimes \dots\otimes V_n$ and
$V'=V_1'\otimes \dots\otimes V_n'$ such that \be\label{Z}
\left(V'_iU_iV_i^{\dagger}\right) Z
\left(V'_iU_iV_i^{\dagger}\right)^{\dagger} = Z\ee for
every $i=1, \dots, n$. Defining \be\label{notations}
D_i&:=& V_i' U_i V_i^{\dagger}\quad (i=1, \dots,
n),\nonumber\\ D&:=& D_1\otimes\dots\otimes D_n,\nonumber\\
|\phi\rangle &:=&V|\psi\rangle\nonumber\\
|\phi'\rangle&:=&V'|\psi\rangle,\ee it follows that $D
|\phi\rangle = |\phi'\rangle$. Note that (\ref{Z}) is
equivalent to $[D_i, Z]=0$ and therefore every $D_i$ is a
diagonal unitary matrix. The operator $D$ will be called a
DLU operator (on $n$ qubits), short for \emph{diagonal
local unitary}. We thus have:

\begin{cor}
    Assume $U=U_1\otimes\dots\otimes U_n$ maps a stabilizer state to a
    stabilizer state. Then, up to the action of local Clifford
    operations, all $U_i$ are diagonal matrices.
\end{cor}


\begin{cor}{\bf (Reduction to diagonal unitaries)}
 \label{cor:diagonalsMain}
 The LU-LC conjecture holds if and only if any two stabilizer states
 that can be mapped onto each other by means of a \emph{diagonal}
 local unitary, are LC equivalent.
\end{cor}

\section{From diagonal LU operations to quadratic forms over GF(2)}
\label{sec:quadraticForms}

Letting $|\psi\rangle$ be an arbitrary stabilizer state, we
consider the expansion \be |\psi\rangle =
\sum_{x\in\mathbb{F}_2^n} \langle x|\psi\rangle\cdot
|x\rangle\ee in the computational basis.
We have used
the standard shorthand notation $|x\rangle =
\bigotimes_{i=1}^n|x_i\rangle$, for every
$x\in\mathbb{F}_2^n$. In this section we will consider the
connection between the components $\langle x|\psi\rangle$
of a stabilizer state and quadratic forms over
$\mathbb{F}_2$. First we introduce some definitions.

Let $m\in\mathbb{N}_0$. A function
$q:\mathbb{F}_2^m\to\mathbb{F}_2$ is called a quadratic
form if there exist coefficients
$\theta_{ij}\in\mathbb{F}_2$ ($i, j=1, \dots, m$, $i<j$)
and a vector $\lambda\in\mathbb{F}_2^m$ such that
\be\label{quadr} q(x) = \sum_{i<j} \theta_{ij}x_ix_j +
\lambda^Tx\ee for every $x=(x_1, \dots,
x_m)\in\mathbb{F}_2^m$. The first term in the r.h.s. of
(\ref{quadr}) is called the quadratic part of the representation of $q$ and the
second term is called its linear part.

We also need some definitions regarding affine spaces over
$\mathbb{F}_2$. Let $S$ be a $k$-dimensional subspace of
$\mathbb{F}_2^n$.
Letting $t$ be a vector in
$\mathbb{F}_2^n$, the \emph{affine space} with \emph{directional
vector space} $S$ and \emph{base point} $t$ is the
set \be S+t := \{y+t\ |\ y\in S\}.\ee
We can now state the connection
between quadratic forms and stabilizer states by recalling
the following result of Ref. \cite{stab_clif_GF2}.

\begin{thm}\cite{stab_clif_GF2}\label{jeroen}
Let $|\psi\rangle$ be a stabilizer state on $n$ qubits.
Then there exist
\begin{itemize}
\item[(i)] a linear subspace $S$ of $\mathbb{F}_2^n$,
\item[(ii)] a quadratic form
$q:\mathbb{F}_2^n\to\mathbb{F}_2$, and \item[(iii)] vectors
$d, t\in\mathbb{F}_2^n$,
\end{itemize}
such that \be\label{stabilizerCoefficients}
2^{k/2}\cdot\langle x|\psi\rangle= \left\{\begin{array}{cl}
i^{d^Ty}
(-1)^{q(y)}& \mbox{ for every } x = y + t \mbox{ with }y\in S\\
0& \mbox{ otherwise,}\end{array} \right.\ee where the algebra in the
exponent of the complex number $i$ is to be performed over
$\mathbb{F}_2$ (i.e., modulo 2). Conversely, every state
$|\psi\rangle$ with components $\langle x|\psi\rangle$
satisfying the above conditions, is a stabilizer state.
\end{thm}
Qualitatively, this result states that, first, the nonzero
components $\langle x|\psi\rangle$ can only be equal to
$\pm 1$ or $\pm i$ (up to an overall normalization);
second, the distribution of the $\pm 1$'s and $\pm i$'s is
governed by quadratic and linear forms, respectively;
third, the nonzero components $\langle x|\psi\rangle$ are
organized in such a way that the corresponding vectors $x$
lie in an affine subspace $S+t$ of
$\mathbb{F}_2^n$.

The following lemma shows that only $S$ and $q$ are essential to the
problem at hand. Anticipating this result, we say that a stabilizer
state is \emph{in standard form} if the parameters $d$ and $t$ vanish.

\begin{lem}{\bf (Reduction to $t=d=0$)}\label{DLC_lemma}
    The LU-LC conjecture is true for general stabilizer states if and
    only if any two DLU-equivalent stabilizer states in standard form
    are also LC equivalent.
\end{lem}

{\it Proof: }
    The ``only if'' part is trivial. To prove the ``if'' direction, assume that any two DLU-equivalent
    stabilizer states in standard form are also LC equivalent.

    Let $|\psi\rangle, |\psi'\rangle$ be general stabilizer states and
    let $D$ be a local unitary s.t.\ $D|\psi\rangle=|\psi'\rangle$. By
    Corollary \ref{cor:diagonalsMain}, we can assume that $D$ is
    diagonal. Let $t,d,S$ and $t',d', S'$ be the parameters associated to
    $|\psi\rangle$ and $|\psi'\rangle$ respectively. Note that $S=S'$ and $t=t'$ as $D$ is diagonal.
    In particular, one
    has
    \begin{equation}
        |\psi\rangle=
        \frac{1}{|S|^{1/2}} \sum_{y\in S} i^{d^T y} (-1)^{q(y)}
        |y+t\rangle.
    \end{equation}
  Set $X(t)=X^{t_1}\otimes\dots\otimes
    X^{t_n}$ and likewise $T^{\dagger}(d)=({T^\dagger})^{d_1}\otimes\dots\otimes
    ({T^\dagger})^{d_n}$, where $T=\operatorname{diag}(1,i)$ is the phase
    gate. One then finds that
    \begin{equation}\label{SF}
        |\psi\rangle_{\text{SF}}:=T^{\dagger}(d)\,X(t)\,|\psi\rangle=
        \frac{1}{|S|^{1/2}} \sum_{y\in S} (-1)^{\tilde q(y)} |y\rangle,
    \end{equation}
    where we have used the notation \be\tilde q(y) = q(y) + \sum_{k<j}
    d_ky_kd_jy_j.\ee In order to prove (\ref{SF}), one uses that $i^ai^b =
    i^{a+b}(-1)^{ab}$ for every $a, b\in\mathbb{F}_2$, where the exponent of $i$ is computed over $\mathbb{F}_2$.
    One therefore has \be i^{d^Ty} = \left\{\prod_{j=1}^n
i^{d_jy_j}\right\} (-1)^{\sum_{k<j}
    d_ky_kd_jy_j}.\ee
     Note that $|\psi\rangle_{\text{SF}}$ is in standard form. The same is true for
    $|\psi'\rangle_{\text{SF}}:=T^{\dagger}(d')\,X(t)|\psi'\rangle$. As a consequence, the local unitary
    operator
    \begin{equation}
        D_{\text{SF}}=X(t) T(d')\,D\,T(d)X(t)
    \end{equation}
    maps $|\psi\rangle_{\text{SF}}$ to $|\psi'\rangle_{\text{SF}}$.
    Because $X$ sends diagonal operators to diagonal operators under
    conjugation, the standard form states are even DLU equivalent.
    Invoking the initial assumption, we conclude that $D_{\text{SF}}$ can be substituted by an LC
    operation. As $X$ and $T$ are Clifford operations, this
    implies that $D$ can be replaced by an LC operation.
\finpr

Now assume that $|\psi\rangle$ is a stabilizer state. Let
\begin{equation}
    D=\left[\begin{array}{cc} 1 & \\ & c_1\end{array}
    \right]\otimes\dots\otimes \left[\begin{array}{cc} 1 & \\ &
    c_n\end{array} \right]
\end{equation}
be a DLU operation defined by the complex phases $c_i$.
The operator $D$ is Clifford if and only if all $c_i\in\{\pm1, \pm
i\}$.
Suppose that $|\psi'\rangle:=D|\psi\rangle$ is again a stabilizer
state. In accordance with Lemma \ref{DLC_lemma}, we take
$|\psi\rangle, |\psi'\rangle$ of the form
\begin{eqnarray*}
    |\psi\rangle&=&
    \frac{1}{|S|^{1/2}} \sum_{x\in S} (-1)^{q(x)} |x\rangle,\\
    |\psi'\rangle&=&
    \frac{1}{|S|^{1/2}} \sum_{x\in S} (-1)^{q'(x)} |x\rangle.
\end{eqnarray*}
Evaluating the equation $\langle x|D|\psi\rangle=\langle
x|\psi'\rangle$, we find for all $x\in S$
\begin{equation}\label{eqn:linearQuadratic}
    \prod_i c_i^{x_i} = (-1)^{q(x)+q'(x)}=(-1)^{Q(x)},
\end{equation}
where we have set $Q(x)=q(x)+q'(x)$. Note that $Q(x)$ is again a
quadratic form and, conversely, every quadratic form can occur this
way.

Equation (\ref{eqn:linearQuadratic}) has an interesting
structure. The l.h.s.\ of this equation has the structure
of an exponentiated \emph{complex linear form}; writing
$c_j:= e^{i\theta_j}$,  one has \be x\to
e^{i(\theta_1x_1+\dots+ \theta_nx_n)}.\ee On the other
hand, the r.h.s. of (\ref{eqn:linearQuadratic}) is an
exponentiated \emph{quadratic form over GF(2)}: \be x\to
(-1)^{Q(x)}.\ee Can one use complex linear mappings to
emulate the behavior of a quadratic form? If the vector
space $S$ is too large, this is clearly impossible. Assume,
e.g., that $e_i$, the $i$th canonical basis vector of
$\mathbb{F}_2^n$, is an element of $S$. Then the r.h.s.\ of
(\ref{eqn:linearQuadratic}) evaluated on $e_i$ gives $c_i$,
which can be of the form $(-1)^{Q(e_i)}$ only if $c_i\in
\{\pm1\}$. Perhaps surprisingly, it turns out that for some
vector spaces $S$, one can represent non-trivial quadratic
forms using complex phases $c_i$. Here is one example:
\begin{equation*}
    S=\left\{
    s_0=
    \left(
        \begin{array}{c}
            0\\0\\0
        \end{array}
    \right),
    s_1=\left(
        \begin{array}{c}
            1\\1\\0
        \end{array}
    \right),
    s_2=\left(
        \begin{array}{c}
            0\\1\\1
        \end{array}
    \right),
    s_3=
    \left(
        \begin{array}{c}
            1\\0\\1
        \end{array}
    \right)
    \right\}\subset \mathbb{F}_2^3,\\
\end{equation*}
and
\begin{equation*}
    c_1=c_2=i,\quad c_3=-i.
\end{equation*}
Then $f: x\mapsto \prod_i c_i^{x_i}$ gives
\begin{equation*}
    f(s_0)=1,\quad f(s_1)=-1,\quad f(s_2)=1,\quad f(s_3)=1.
\end{equation*}
One can easily check that $f$ represents a quadratic form on $S$.
Also, it is impossible to realize $f$ by means of phases
$c_i\in\{\pm1\}$ (that is because any set of real phases would give
rise to an even number of $-1$'s, whereas $f$ is negative only once).

Hence sometimes it does pay off to leave the set of real
phases in (\ref{eqn:linearQuadratic}), even if one aims to
represent a form which takes on values only in $\{\pm1\}$.
The preceding example is no threat to the LU-LC conjecture,
as we only had to go to fourth roots of unity and $c_i \in
\{\pm 1, \pm i\}$ still induce Clifford operations, as in
this case the matrix \be \bigotimes_i \left[ \begin{array}{cc} 1&0\\
0& c_i\end{array}\right]\ee is still a local Clifford
operation. The LU-LC conjecture amounts to claiming that it
is never necessary to go to more general phases when
representing quadratic forms over GF(2) by way of
(\ref{eqn:linearQuadratic}).

\begin{thm}{\bf (Reduction to quadratic forms)}
    Let $S$ be a linear subspace of $\mathbb{F}_2^n$, and let
    $Q:\mathbb{F}_2^n\to\mathbb{F}_2$ be a quadratic function. Suppose
    that there exist complex phases $c_1, \dots, c_n$, (i.e.\ $c_i$ is a
    complex number of modulus one) such that
    \be
        (-1)^{Q(x)} = \prod_{i=1}^n c_i^{x_i},\quad  \mbox{for every $x\in S$}.
    \ee
    If, for every such $Q$ and $S$,  the phases $c_i$ can always be chosen from $\{\pm1, \pm i\}$,
    then the LU-LC conjecture is true.

    Conversely, assume the LU-LC conjecture holds. Additionally, assume
    that if two stabilizer states can be mapped onto each other by means
    of a diagonal local unitary, then also by a {\bf diagonal} local
    Clifford operation. Then the phases $c_i$ introduced above
    can always be chosen from the set $\{\pm1, \pm i\}$.
\end{thm}

\begin{proof}
    Immediate from the preceding discussion.
\end{proof}

\section{From stabilizer codes to stabilizer states}
\label{sec:fromCodesToStates}

In this section, we prove that the LU-LC conjectures for
stabilizer codes and stabilizer states are equivalent.
Section \ref{sect_prelim_codes_to_states} introduces some
additional preliminary results regarding stabilizer codes.
The proof is given in Section
\ref{sect_proof_codes_to_states}. The intuition behind the
argument is to assign to a code $\rho$ on $n$ qubits a
\emph{purification} $\sigma=|\Psi\rangle\langle\Psi|$; more
concretely, we will extend the qubits $\{1,\dots,n\}=:A$ by
auxiliary systems $\{n+1,\dots,n+l\}=:B$ and define a
stabilizer state $\ket\Psi$ on the extended space (i.e., on
$n+l$ qubits) in such a way that
$\tr_B\ket\Psi\bra\Psi=\rho$.  For suitable choices of the
purifications for the LU equivalent codes $\rho$ and
$\rho'$, we find that the LU equivalence of these codes
implies the LU equivalence of their purifications. We then
invoke the assumption that the LU-LC conjecture for
stabilizer states is correct, implying that the
purifications are actually LC equivalent. Finally, it is an
easy step to prove that the LC equivalence of the
purifications implies LC equivalence of the codes $\rho$
and $\rho'$.

\subsection{Preliminaries}\label{sect_prelim_codes_to_states}

An important feature of stabilizer states and codes is that
they allow for an efficient description in terms of
subspaces of the binary vector space $\mathbb{F}_2^{2n}$,
as will be made explicit next. We refer to Refs.
\cite{Go97, QCQI, Gr05} for more details.

First, the connection between binary vector spaces and
Pauli operators is provided by the map ${\mathcal
W}:\mathbb{F}_2^{2n}\to {\mathcal G}_n$ defined by
\begin{equation}
    {\mathcal W}(z,x) =\bigotimes_{i=1}^n i^{z_i x_i}
    Z^{z_i} X^{x_i},
\end{equation}
where $z=(z_1, \dots, z_n),\  x=(x_1, \dots, x_n) \in
\mathbb{F}_2^{n}$. For a vector $v=(z, x)\in
\mathbb{F}_2^{2n}$, we set ${\mathcal W}(v):={\mathcal
W}(z,x)$. Invoking (\ref{eqn:shortXZ}), we have that ${\mathcal
W}(t,0)=Z(t)$ and ${\mathcal W}(0,t)=X(t)$, for every
$t\in\mathbb{F}_2^n$.

It can be checked by direct computation that
two Pauli operators ${\mathcal W}(z,x)$ and ${\mathcal
W}(z',x')$ commute if and only if
\begin{equation}
    [
        \left(\begin{array}{c}z\\x\end{array}\right),
        \left(\begin{array}{c}z'\\x'\end{array}\right)
    ]:=z^T x'+x^T z'=0.
\end{equation}
The square bracket will be referred to as the \emph{symplectic
inner product} of the binary vectors $(z,x)$ and $(z',x')$.

We now consider a $k$-dimensional linear subspace $M$ of
$\mathbb{F}_2^{2n}$, where a basis $\{m^{(1)},\dots,m^{(k)}\}$ has been
chosen. We further assume that $M$ is an \emph{isotropic
subspace}, i.e., the symplectic inner product between any two
vectors in $M$ vanishes. Lastly, we choose a vector $v=(v_1,
\dots, v_k) \in \mathbb{F}_2^k$ and consider the set
\begin{equation}\label{set}
    \{(-1)^{v_1} {\mathcal W}(m^{(1)}), \dots, (-1)^{v_k} {\mathcal W}(m^{(k)})\}.
\end{equation}
One can then verify that the multiplicative group ${\mathcal S}$
generated by the elements in the set (\ref{set}) is a stabilizer
of rank $k$. Conversely, it is well known that any stabilizer can
be obtained by means of the above construction (see e.g. \cite{QCQI}).

The basis vectors $\{m^{(i)}\}$ of the $k$-dimensional
isotropic subspace $M\subseteq\mathbb{F}_2^{2n}$ are
usually arranged as the columns of a $2n\times k$ matrix
$R$ over $\mathbb{F}_2$, which is said to be a
\emph{generator matrix} associated with the space $M$.

Next we state two lemmas that will be used below. To do this, we
need some additional notations: let $R$ be a generator matrix of a
$k-$dimensional isotropic subspace $M\subseteq\mathbb{F}_2^{2n}$.
Given a set of vectors $v_1,\dots,v_l$ in $\mathbb{F}_2^{2n}$, we
denote by $\left[R,v_1,\dots,v_l\right]$ the $2n\times(k+l)$
matrix obtained by appending the vectors $v_i$ as further columns
to $R$ (this notation involving square brackets is not to be
confused with the notation for the symplectic inner product).

Qualitatively, the next lemma shows that one can complete any
stabilizer group ${\mathcal S}$ to a maximal one of order $2^n$ by
adding suitable ``$Z$-type'' operators.

\begin{lem}
\label{extensionLemma}
    Let $\rho$ be a stabilizer code on $n$ qubits. Let ${\mathcal S}$ be its stabilizer, let $R$
    be an associated generator matrix, and let $k$ be the rank of ${\mathcal S}$.
    Then there exist vectors $z^{(1)}, \dots, z^{(n-k)} \in \mathbb{F}_2^n$
such that
    \be
        \left[R,\left(z^{(1)}\atop0\right),\dots,\left(z^{(n-k)}\atop0\right)\right]
    \ee
    is a generator matrix of a stabilizer state on $n$ qubits.
\end{lem}

{\it Proof:}
    Let $R$ be a $2n\times k$ generator matrix of $M$. We can always choose $R$ such
    that its lower $n\times k$ submatrix consists of $k'$ linearly
    independent columns followed by $k-k'$ columns containing only
    zeros, for some $k'\leq k$. So
    \be
        R=
        \left[
            \begin{array}{cc}
                P_1 & P_2 \\
                Q_1 & 0
            \end{array}
        \right]
    \ee
    where $P_1$ and $Q_1$ are $n\times k'$ matrices and $P_2$ has dimensions $n\times
    (k-k')$; also, $Q_1$ and $P_2$ have full rank.
    Consider the orthogonal complement of the column space of $Q_1$, denoted in a shorthand notation by
    $\langle Q_1\rangle^\bot$.
    This space has dimension $n-k'$ and contains the column space of $P_2$ as a
    $k-k'$ dimensional subspace; this follows from the property that $M$ is isotropic.
    Hence, there exists an $n\times (n-k)$ matrix $P_3$ such that $[P_2 \ P_3]$
    is an $n\times (n-k')$ generator matrix of $\langle Q_1\rangle^\bot$. It then follows that
    \be
        \left[
            \begin{array}{ccc}
                P_1 & P_2 & P_3 \\
                Q_1 & 0 & 0
            \end{array}
        \right]
    \ee
    is a $2n\times n$ generator matrix of an $n$-dimensional isotropic space.
    This proves the result.\finpr

The following lemma is taken from the standard reference
\cite{QCQI}.

\begin{lem}\cite{QCQI}
    \label{hyperbolicCouples}
Let ${\mathcal S}$ be a stabilizer on $n$ qubits generated by $k$
independent elements $g_1, \dots, g_{k}$. Let $i$ be any fixed
number in the range $1, \dots, k$. Then there exists $g\in{\mathcal
G}_n$ such that $gg_i=-g_ig$ and $gg_j=g_jg$ for every $j=1,
\dots, k$, $j\neq i$.
\end{lem}

\subsection{Reduction to stabilizer states}
\label{sect_proof_codes_to_states}

We proceed to the proof of Theorem \ref{thm:statesSuffice}. As
stabilizer states are contained in the set of stabilizer codes, the
non-trivial part of the theorem is: if the LU-LC conjecture is true
for states, then also for codes. So for the rest of this section, we
assume validity of the LU-LC conjecture for states.

First we look for a suitable purification of $\rho$. Let $\rho$ be
a rank $k$ stabilizer code on $n$ qubits with stabilizer ${\mathcal S}$, let $\{z^{(1)}, \dots, z^{(n-k)}\}$ be
as in Lemma \ref{extensionLemma}, and set $l:=n-k$. For every $y\in
\mathbb{F}_2^{l}$, let ${\mathcal S}_y$ be the stabilizer generated by the set of operators
\begin{equation}\label{swhy}
    \left\{{\mathcal S},\ (-1)^{y_1}Z(z^{(1)}), \dots, (-1)^{y_{l}}Z(z^{(l)})
    \right\}
\end{equation}
and let $\ket{\psi_y}$ be the stabilizer state on $n$ qubits with
stabilizer ${\mathcal S}_y$. The $(n+l)$-qubit state \be
\ket\Psi:=\sum_{y\in\mathbb{F}_2^l}\ket{\psi_y}\otimes\ket y \ee
will be our candidate for a purification of the state $\rho$.
Therefore, we need to prove that \begin{itemize}\item[(i)]
$\ket\Psi$ is a stabilizer state, and \item[(ii)] the partial
trace of $|\Psi\rangle\bra\Psi$ over the qubits in $\{n+1,
\dots, n+l\}$ is
 equal to the state $\rho$.\end{itemize} These statements
are proven next.

To prove (i), we will construct a maximal stabilizer on $n+l$
qubits having the state $\ket\Psi$ as a fixed point. First, let
$\{g^{(1)}, \dots, g^{(k)}\}$ be a generating set of ${\mathcal S}$. It can
then easily be verified that \be g^{(i)}\otimes
I_l\ket\Psi=\ket\Psi\ee for every $i=1, \dots, k$. Second, for
every $j=1, \dots, l$, the calculation \be Z(z^{(j)}, e^{(j)})
\ket\Psi&=& \sum_y Z(z^{(j)})\ket{\psi_y}\otimes Z(e^{(j)})\ket y
\nonumber\\&=& \sum_y (-1)^{y_j}\ket{\psi_y}\otimes (-1)^{y_j}\ket
y\nonumber \\ &=& \ket\Psi \ee shows that the operators $Z(z^{(j)},
e^{(j)})$ also fix the state $\ket\Psi$. Here, $e^{(j)}$ is the $j$th
canonical basis vector of $\mathbb{F}_2^l$. Finally, it follows
from Lemma \ref{hyperbolicCouples} that there exist $l$ Pauli
operators $h^{(1)}, \dots, h^{(l)}\in{\mathcal G}_n$ such that \be
h^{(j)}|\psi_y\rangle = |\psi_{y+e^{(j)}}\rangle,\ee for every $j=1,
\dots, l$ and $y\in\mathbb{F}_2^l$. We then have \be h^{(j)}\otimes
X(e^{(j)}) \ket\Psi&=& \sum_y h^{(j)}\ket{\psi_y}\otimes X(e^{(j)})\ket
y \nonumber\\&=& \sum_y \ket{\psi_{y+e^{(j)}}}\otimes \ket{y+e^{(j)}} =
\ket\Psi.\ee Thus, all $n+ l$ operators in the set \be
\left\{g^{(i)}\otimes I_l,\ Z(z^{(j)}, e^{(j)}),\ h^{(j)}\otimes
X(e^{(j)})\right\}_{i,j},\ee where $i=1, \dots, k$ and $j=1, \dots, l$,
stabilize the state $\ket\Psi$. Moreover, these operators generate
a rank $n+l$ stabilizer,
showing that $\ket\Psi$ is indeed a stabilizer state.

We now prove (ii). The kets $\ket{\psi_y}$ form a basis within the
range of $\rho$. To see this, recall that any two stabilizer
states whose stabilizer operators differ only by global phases are
orthogonal. Thus, $\{\ket{\psi_y}\}$ is a set of $2^{n-k}$
mutually orthogonal states, all of which stabilized by any
$g\in {\mathcal S}$. Further, all these states are eigenvectors of
$\rho$ with eigenvalue $|{\mathcal S}| = 2^{k-n}$. But the rank of
$\rho$ is equal to $2^{n-k}$ as well, and therefore
\be
    \rho = 2^{k-n}\sum_y \ket{\psi_y}\bra{\psi_y}.
\ee

We are now in a position to prove the main result of this
section. Let $\rho$, $U$, ${\mathcal S}$, $\ket\Psi$ and
$\{\ket{\psi_y}\}$ be as above. By the same reasoning as
the one employed in Section \ref{sec:fromLUtoDLU}, there is
no loss of generality in assuming that $U$ is diagonal. Set
$\rho'=U\rho U^\dagger$ and let ${\mathcal S}$ be the
stabilizer of $\rho'$.

        First we claim that $\ket{\psi_y'}:=U\ket{\psi_y}$ is a stabilizer
        state, for each $y\in\mathbb{F}_2^l$.  Indeed, it follows from
        $\rho\ket{\psi_y}=\ket{\psi_y}$ and the definition of $\rho'$ that
        $\ket{\psi_y'}$ is an eigenvector of $\rho'$ with eigenvalue 1,
        and hence of each $g'\in {\mathcal S}'$. Further, by construction,
        we have \be Z{(z^{(j)})}\ket{\psi_y}=(-1)^{y_j}\ket{\psi_y}\ee and
        hence \be U Z{(z^{(j)})}
        U^\dagger\ket{\psi_y'}=(-1)^{y_j}\ket{\psi_y'},\ee for every $j=1,
        \dots, l$. As $U$ is diagonal, it commutes with $Z{(z^{(j)})}$, which
        finally implies that $\ket{\psi_y'}$ is an eigenvector of
        $Z{(z^{(j)})}$ with the eigenvalue $(-1)^{y_i}$.  Define ${\mathcal
        S}_y'$ by substituting ${\mathcal S}$ by ${\mathcal S}'$ in
        (\ref{swhy}).  The group ${\mathcal S}_y'$ can be checked to be a
        stabilizer of rank $n$ with $\ket{\psi_y'}$ as a common
        eigenvector. This shows that $\ket{\psi_y'}$ is a stabilizer state
        with stabilizer ${\mathcal S}_y$, for every $y\in\mathbb{F}_2^l$.

    It now follows from an analogous argument as made in the beginning of this section that the state
    \be
        |\Psi'\rangle:= \sum_y \ket{\psi_y'}\otimes \ket y
    \ee
    is a stabilizer state on $n+l$ qubits such that $\rho'$ is equal
    to the partial trace of this state over the qubits in the set $\{n+1, \dots, n+l\}$. Furthermore,
    by definition of the states $|\psi_y'\rangle $ one
    has \be(U\otimes I_l) \ket\Psi = |\Psi'\rangle,\ee i.e., the states $|\Psi'\rangle$ and $|\Psi\rangle$ are LU equivalent.
    Assuming validity of the LU-LC conjecture,
        there
    exists a  LC operator on $n+l$ qubits relating these two states. Taking
    the partial trace over the qubits in the set $\{n+1, \dots, n+l\}$
    then shows that $\rho$ and $\rho'$ are LC equivalent.
    This proves Theorem \ref{thm:statesSuffice}.

\section{Outlook}

Unfortunately, even the strong reductions presented in this paper did
not suffice to resolve the LU-LC conjecture. There are, however,
further routes which may merit exploration. For example, we have
indications for the fact that the phases $c_i$ appearing in Theorem
\ref{thm:quadraticForms} may always be taken to be roots of unities
(i.e. of the form $e^{i\pi\phi}$, for $\phi\in\mathbb{Q}$). This can
be shown to imply that each $c_i$ is a power of $e^{2\pi i / 2^l}$ for
some $l$ and the LU-LC problem would reduce to a statement concerning
the solutions of certain systems of linear equations in modules over
the ring $\mathbb{Z}_{2^l}$. We did not make these arguments explicit,
as even employing this additional structure, a general solution
remains elusive.

\section{Acknowledgements}

DG is pleased to acknowledge the support of Jens Eisert during many
stages of this project.

This work has been supported by the European Union (QICS, OLAQUI,
SCALA, QAP, EURYI grant of J.\ Eisert), the FWF, and the EPSRC (IRC
QIC).


\end{document}